\documentclass{IEEE_lsens}

\usepackage{cite}
\usepackage{graphicx}
\usepackage{booktabs}
\usepackage{array}
\usepackage{url}
\usepackage{textcomp}

\graphicspath{{./}}

\renewcommand{\thetable}{\Roman{table}}

\begin{document}

\markboth{IEEE Sensors Letters}{Swami and Sonawane: Wire-Level Interrupt-to-Decision Latency of On-Sensor MLC versus Host Inference}

\IEEELSENSarticlesubject{Sensor System Integration, Intelligent Sensing}

\title{Wire-Level Interrupt-to-Decision Latency of On-Sensor MLC versus Host Inference on the NVIDIA Jetson Orin Nano: A Pre-Registered Measurement Study}

\author{\IEEEauthorblockN{Akul~Swami\IEEEauthorrefmark{1} and Dnyaneshwar~Sonawane\IEEEauthorrefmark{2}}
\IEEEauthorblockA{\IEEEauthorrefmark{1}Independent Researcher (e-mail: swami.akul@alumni.uml.edu)\\
\IEEEauthorrefmark{2}Independent Researcher (e-mail: sonawane.dnyaneshwar19@gmail.com)}
\thanks{Corresponding author: A. Swami (e-mail: swami.akul@alumni.uml.edu).}%
\thanks{Code, data, and the full pre-registration chain are available at https://github.com/akulswami/sensor-mlc-latency.}%
\thanks{This work has been submitted to the IEEE for possible publication. Copyright may be transferred without notice, after which this version may no longer be accessible.}
}

\IEEELSENSmanuscriptreceived{Manuscript submitted May 28, 2026.}

\IEEEtitleabstractindextext{%
\begin{abstract}
The Machine Learning Core (MLC) embedded in the STMicroelectronics LSM6DSOX IMU is widely cited as a low-latency alternative to host-side inference, yet wire-level decision-delivery latency is rarely measured. Using a Saleae Logic Pro 8 logic analyzer on an NVIDIA Jetson Orin Nano, we measured interrupt-to-decision latency (sensor INT1 edge to host decision GPIO) for three pipelines (a host-side decision-tree classifier, the standard MLC bank-switch read protocol, and an MLC binary-fast variant) under idle, I\textsuperscript{2}C bus contention, and CPU stress. The protocol was pre-registered with 13 externally-timestamped Zenodo amendments (4,770 of 4,860 trials included, 98.15\%, across nine cells). The host pipeline exhibits lower median latency than the MLC pipeline under all conditions: 321.7 vs 681.5 $\mu$s at idle (2.1$\times$ faster) and 574.5 vs 1,325.4 $\mu$s under I\textsuperscript{2}C contention (2.3$\times$ faster). The three-transaction I\textsuperscript{2}C read protocol, not the silicon's classification, is the dominant latency contributor. We additionally characterize a reproducible 706.5 ms MLC decision cadence that bounds full stimulus-to-decision latency. Code, data, and pre-registration: github.com/akulswami/sensor-mlc-latency.
\end{abstract}

\begin{IEEEkeywords}
Inertial measurement unit, machine learning core, edge inference, latency measurement, pre-registration, LSM6DSOX.
\end{IEEEkeywords}}

\maketitle

\section{Introduction}

Inertial measurement units (IMUs) with embedded Machine Learning Cores (MLCs) are increasingly deployed in edge applications spanning wearables, industrial monitoring, and biomedical instrumentation \cite{ref1}. The MLC paradigm runs a decision-tree classifier inside the sensor and exposes only an output register to the host. It is often framed as a latency and energy advantage over host-side classification \cite{ref2,ref3}, on the implicit assumption that on-sensor inference is unconditionally lower-latency because the host need only poll a single register.

This assumption is rarely tested with wire-level instrumentation. Vendor application notes characterize MLC architecture and per-class accuracy but not end-to-end stimulus-to-decision latency \cite{ref2}. Even recent safety-critical applications such as exoskeleton control \cite{ref3} cite the on-sensor latency advantage without measuring it directly. The resulting gap between ``the classification is correct'' and ``the classification reaches the host in time to act'' is the failure mode conventional functional testing misses. Framed in sensor-system terms, the question is whether a commercial on-sensor inference engine can deliver its decision to the host within a latency budget, a deployment-capability property of the sensor device itself, dictated by how it exposes its on-chip inference output, not by any host-side algorithm.

We measure wire-level interrupt-to-decision latency (sensor INT1 edge to host decision GPIO) for three pipelines on a representative edge platform (NVIDIA Jetson Orin Nano + STMicroelectronics LSM6DSOX over I\textsuperscript{2}C) using a Saleae Logic Pro 8 logic analyzer: (a) a host-side decision-tree classifier reading raw accelerometer data, (b) the standard MLC bank-switch read protocol, and (c) an MLC binary-fast variant that omits the I\textsuperscript{2}C read entirely, toggling the decision GPIO on every MLC interrupt (valid for a 2-class configuration). Measurements span idle, I\textsuperscript{2}C bus contention, and CPU saturation. The full protocol was pre-registered with 13 externally-timestamped Zenodo amendments \cite{ref4}.

The principal finding is that the host pipeline exhibits lower median wire-level latency than the standard MLC bank-switch pipeline under all tested conditions, with the three-transaction I\textsuperscript{2}C read protocol, not the silicon's classification, as the dominant latency contributor. Thus, the contribution is a quantitative characterization of a sensor-device output path (INT1 assertion, embedded-bank readout, and host-visible decision delivery), not a new AI/ML classifier. Our contributions are: (1) the first pre-registered wire-level latency characterization of the LSM6DSOX MLC versus host-side inference (4,770 of 4,860 trials included, 98.15\%, across nine pipeline$\times$condition cells); (2) empirical identification of the LSM6DSOX MLC's mandatory embedded-bank read protocol, not its classifier, as the dominant latency contributor, a property intrinsic to the sensor's output architecture; (3) characterization of a reproducible 706.5 ms MLC decision cadence (the 75-sample $\times$ 104 Hz window period) that bounds full stimulus-to-decision latency; and (4) pre-registration with externally-timestamped Zenodo DOIs as an audit-defensible framework for sensor-measurement claims.

\section{Background and Related Work}

\subsection{On-sensor inference in MEMS IMUs}

Embedded machine learning in 6-axis microelectromechanical systems (MEMS) IMUs is now a standard product feature. STMicroelectronics introduced the MLC in the LSM6DSOX \cite{ref2}, a configurable decision-tree engine running at the accelerometer ODR (104 Hz in our deployment) that consumes features over sliding sample windows and emits class labels the host reads via I\textsuperscript{2}C bank-switched registers. Successor parts (LSM6DSV16X, LSM6DSO32X) retain this architecture, and competing vendors offer analogues (Bosch BHI260AP, Analog Devices ADXL367). The consistent commercial messaging is that on-sensor inference reduces data-bus traffic, host CPU utilization, and end-to-end latency. The first two claims are testable from system-level metrics; the third is testable only with wire-level instrumentation, which is rarely performed.

\subsection{Wire-level latency in safety-critical edge ML}

Safety-critical edge applications require bounded stimulus-to-actuation latency, not merely correct classification. Existing MLC documentation emphasizes architecture and accuracy \cite{ref2}, while independent characterizations rarely isolate the sensor read protocol's contribution to decision-delivery delay.

\subsection{Pre-registration in sensor measurement}

Pre-registration is used here to separate pre-specified sensor-measurement hypotheses from post-hoc interpretation.

\section{Experimental Setup}

\subsection{Hardware Platform}

The edge platform is an NVIDIA Jetson Orin Nano Developer Kit (JetPack 6.2.2, kernel 5.15.148-tegra, 6-core ARM Cortex-A78AE). All measurements ran under a custom MAXN\_SUPER\_JC nvpmodel (v7.6 \cite{ref4}) pinning all six CPUs at 1728 MHz; per-block jetson\_clocks effectiveness (tegrastats samples $\geq$ 1700 MHz) was $\geq$ 99\%, with all 81 blocks at 100\%.

The sensor is an STMicroelectronics LSM6DSOX 6-axis IMU on an Adafruit breakout (PID 4438; onboard 3.3 V regulation and 10K I\textsuperscript{2}C pull-ups) on I\textsuperscript{2}C bus 7 (pins 3/5, address 0x6A) at 400 kHz (Fast Mode, Jetson default). All three pipelines configure the accelerometer identically (CTRL1\_XL = 0x50: 208 Hz, $\pm$2 g, LPF2 off). The MLC runs at 104 Hz (EMB\_FUNC\_ODR\_CFG\_C = 0x35: MLC\_ODR = 0b11, half the 208 Hz accelerometer ODR) with a custom 2-class motion/still classifier (75-sample windows, config mlc\_motion\_w75.h) trained in ST MEMS Studio, with output routed to INT1.

\begin{figure}[!t]
  \centering
  \includegraphics[width=0.62\columnwidth]{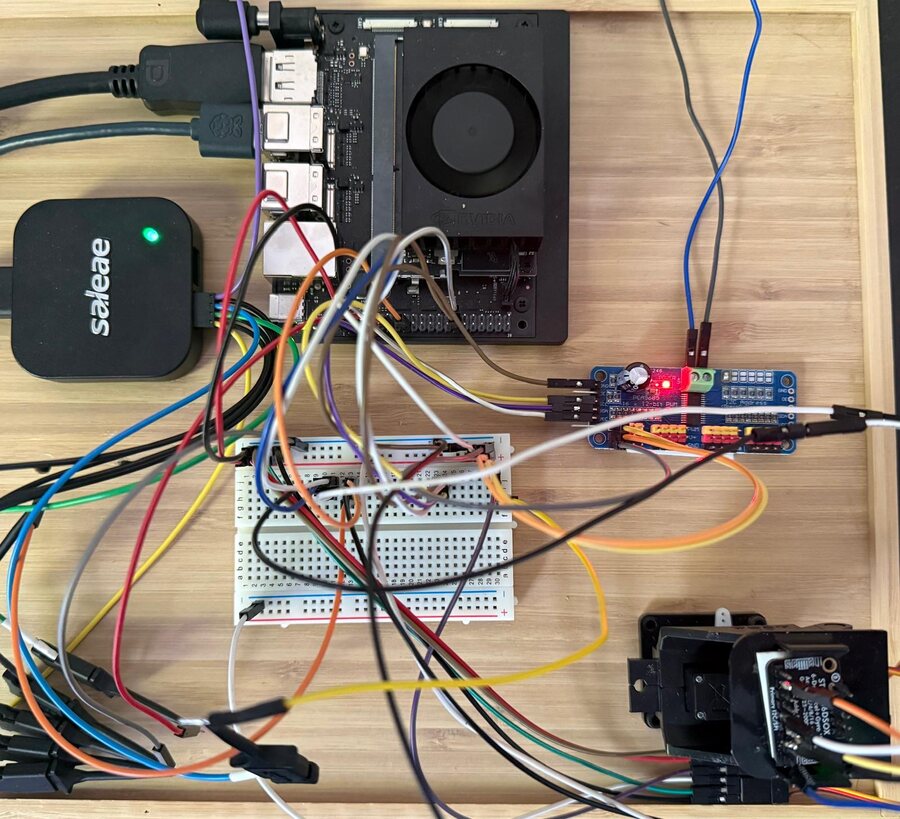}
  \caption{Measurement setup. The LSM6DSOX IMU (on a servo-driven motion stimulus) and the PCA9685 servo controller connect to the NVIDIA Jetson Orin Nano over separate I\textsuperscript{2}C buses; a Saleae Logic Pro 8 taps the sensor INT1 edge (D0), the host decision GPIO (D1), and the servo PWM command (D2) for wire-level timing.}
  \label{fig:setup}
\end{figure}

Two Jetson GPIO lines are instrumented with a Saleae Logic Pro 8 (250 MS/s): \textbf{D0} on pin 15 (INT1 from the LSM6DSOX, gpiochip0 line 85) and \textbf{D1} on pin 11 (decision edge from the pipeline under test, line 112). This yields nanosecond-resolution wire-level timestamps independent of the Jetson clock and of SSH command jitter (both D0 and D1 are wire edges). The servo PWM command is additionally captured on D2 as a timing reference. The wire-level GPIO timestamping methodology follows our prior cross-platform characterization \cite{ref9}. Motion stimulus is an SG90 servo on a PCA9685 PWM controller (I\textsuperscript{2}C bus 1, address 0x60; the board default 0x40 collides with the on-board INA3221, so an address-select pad was bridged to set the A5 bit, relocating the controller to 0x60), commanded over SSH. The full measurement setup is shown in Fig.~\ref{fig:setup}.

\subsection{Pipelines Under Test}

Three pipelines are compared end-to-end, sharing the same I\textsuperscript{2}C arbitration, gpiod write path, and accelerometer configuration, and differing only in what happens after an interrupt arrives:

\textbf{(a) host} (host\_pipeline\_parity): polls the accelerometer at 208 Hz, assembles a 75-sample window, and applies a decision-tree classifier (variance of accelerometer L2-norm against a calibrated threshold), toggling D1 on every output state change.

Host window and MLC window are both 721.2 ms: parity\_core's pc\_step implements a 2:1 decimation of the host sample stream, equalizing the two windows to a common period.

\textbf{(b) mlc} (latency\_test\_mlc\_w75): on each INT1 edge, performs the three I\textsuperscript{2}C transactions of the bank-switch read mandated by the LSM6DSOX register architecture \cite{ref2} (write FUNC\_CFG\_ACCESS = 0x80 to enter the embedded-function bank, read MLC0\_SRC = 0x70, write FUNC\_CFG\_ACCESS = 0x00 to restore the user bank), writing D1 if the output changed.

\textbf{(c) mlc-binary} (latency\_test\_mlc\_binary\_w75): on each INT1 edge, unconditionally toggles D1 without reading MLC0\_SRC, valid for the 2-class case and providing a kernel/gpiod-only latency floor against which (b)'s I\textsuperscript{2}C-read overhead is measured. Because this study measures decision-delivery latency rather than classification accuracy, parity is enforced at the level of identical accelerometer configuration, window length, and binary motion/still decision semantics; the mlc-binary condition isolates the non-classifier read path.

The host pipeline sets INT1\_CTRL = 0x01 (DRDY); mlc and mlc-binary set INT1\_CTRL = 0x00, routing the MLC's decision via MD1\_CFG = 0x02 instead: intrinsic to each pipeline's decision source (raw sample vs. embedded classifier).

\subsection{Stress Conditions}

Three conditions are tested. \textbf{idle}: CPUs pinned but otherwise unloaded. \textbf{i2c-contention}: three concurrent i2c\_hammer processes continuously read non-MLC registers from the LSM6DSOX on bus 7, contending for bus access. \textbf{stress}: stress-ng 0.13.12 saturates all six CPUs (--cpu 6 --cpu-method matrixprod). Hammer-process count and CPU saturation are verified at block start.

\subsection{Measurement Protocol}

Each (pipeline, condition) cell comprises 9 blocks of 300~s. Block order across the 81-block campaign follows a pre-registered deterministic shuffle (seed 1990185399 \cite{ref4}). Within each block the orchestrator delivers 60 candidate transitions; latency is t(D1) $-$ t(D0). Inclusion (pre-registered v7.4) requires exactly one D1 rising edge, preceded by an unambiguous single D0 edge, per stimulus window; violations are categorized and reported as a classifier-stability outcome (Section~V.D). Across the campaign, 4,770 of 4,860 trials (98.15\%) were included, no cell exceeding the pre-registered 10\% exclusion ceiling.

\section{Methodology and Statistical Treatment}

\subsection{Pre-registration}

Hypotheses, exclusions, statistical tests, and multiplicity correction were specified in a public pre-registration chain with Zenodo DOIs \cite{ref4}. The chain records substantive hypothesis and methodology changes before the affected confirmatory data. Three falsified hypotheses remain on the public record: pilot-stage H1 and H4', and confirmatory H7'. Thus, the record distinguishes pre-specified predictions from post-hoc interpretation.

\subsection{Statistical tests and multiplicity correction}

Each hypothesis is matched to a pre-registered test. H1' and H2' (stochastic ordering of two independent samples) use a one-sided Mann-Whitney U test, with the Hodges-Lehmann shift and a 95\% percentile-bootstrap CI (10,000 resamples) as the effect estimate \cite{ref5}, \cite{ref6}. H3' (the interaction contrast $\Delta$\_MLC $-$ $\Delta$\_host \textgreater{} 0) bootstraps the contrast statistic (10,000 resamples). H5' (equivalence within $\pm$30 $\mu$s) uses TOST \cite{ref7}: the 90\% CI on the median difference must lie strictly within [$-$30, +30] $\mu$s. H6' (energy difference \textgreater{} 1,000 mW) uses Mann-Whitney U on VDD\_IN samples with 1/10 subsampling to reduce the autocorrelation of 100 ms-cadence INA3221 readings. H7' (stability ordering) uses a one-sided Fisher's exact test, chosen over chi-square because the per-cell unstable counts (6--15 of 540) fall where the chi-square asymptotic approximation is unreliable.

Multiplicity correction across the {H1', H2', H3', H6', H7'} family is by Holm-Bonferroni at family-wise $\alpha$ = 0.05 \cite{ref8}. H5' uses TOST equivalence rather than a directional p-value and is reported alongside the family rather than within it; this is an analysis-side refinement of the v7.5 specification, which had named a six-hypothesis Holm family.

\section{Results}

\begin{figure}[!t]
  \centering
  \includegraphics[width=\columnwidth]{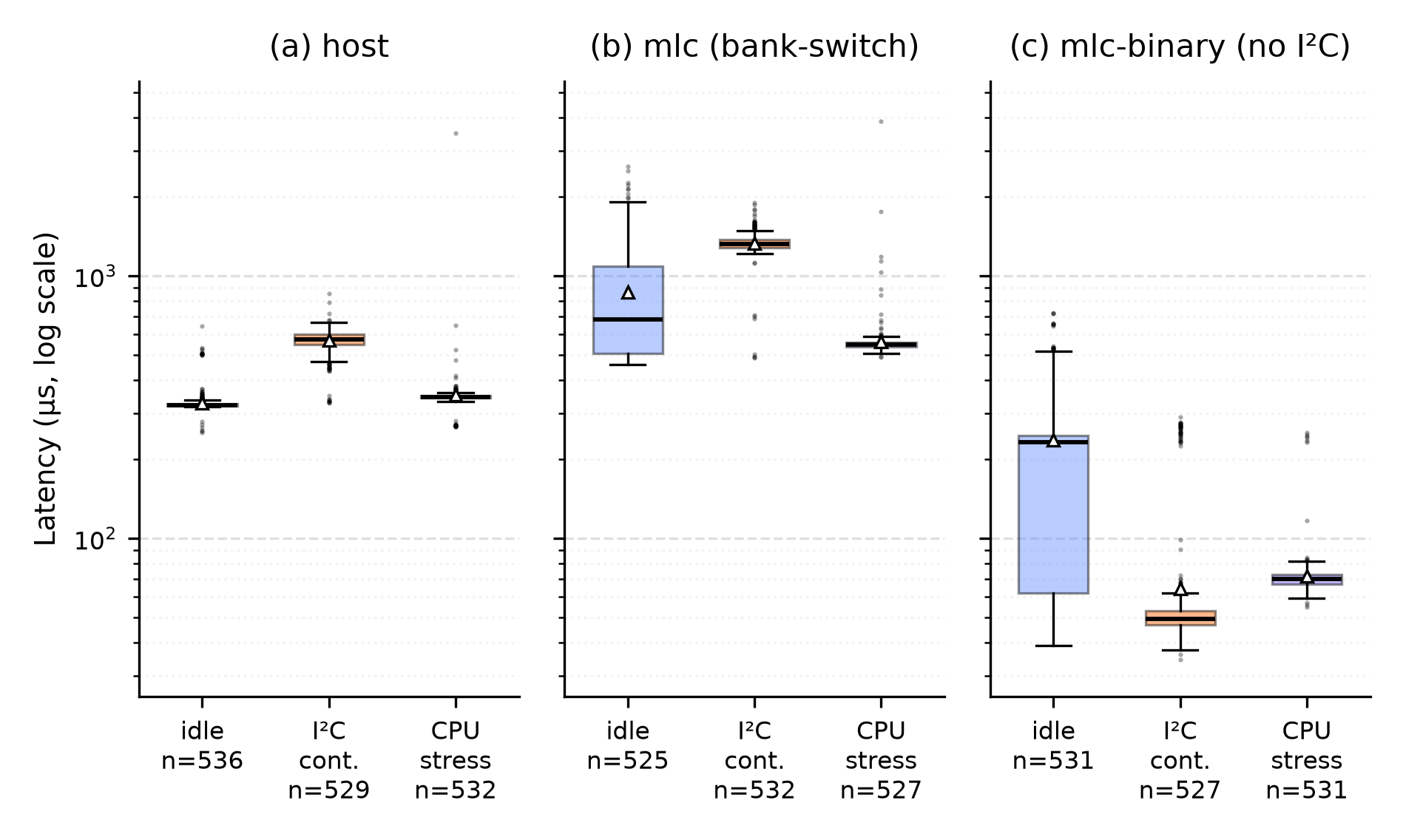}
  \caption{Per-trial wire-level latency by pipeline and condition, on one shared log-scale $y$-axis across all three panels so box position is directly comparable across pipelines. The mlc-binary panel (zero I\textsuperscript{2}C transactions on the decision path) isolates the kernel/gpiod latency floor. Box = IQR (Q1--Q3); whiskers = 1.5$\times$IQR; $\triangle$ = mean; $\bullet$ = outlier; every point is shown in-frame, with no per-panel capping.}
  \label{fig:latency}
\end{figure}

The confirmatory campaign collected 4,860 candidate trials across 81 blocks of 300 s; 4,770 (98.15\%) satisfied the pre-registered inclusion criteria. All blocks achieved 100\% jetson\_clocks effectiveness under MAXN\_SUPER\_JC. Per-cell latency distributions appear in \textbf{Fig.~\ref{fig:latency}}; summary statistics in \textbf{Table~\ref{tab:latency}}.

\begin{table}[!t]
  \centering
  \caption{Wire-Level Latency per Pipeline $\times$ Condition ($\mu$s)}
  \label{tab:latency}
  \setlength{\tabcolsep}{4pt}
  \footnotesize
  \begin{tabular}{llrrr}
    \toprule
    Pipeline & Condition & $n$ & Median ($\mu$s) & IQR ($\mu$s, p25--p75) \\
    \midrule
    host       & idle      & 536 & 321.7  & 319.7--326.3   \\
    host       & i2c-cont. & 529 & 574.5  & 547.8--599.0   \\
    host       & stress    & 532 & 345.0  & 342.0--348.9   \\
    mlc        & idle      & 525 & 681.5  & 505.4--1086.8  \\
    mlc        & i2c-cont. & 532 & 1325.4 & 1283.7--1370.8 \\
    mlc        & stress    & 527 & 546.1  & 535.9--557.0   \\
    mlc-binary & idle      & 531 & 231.9  & 61.8--246.2    \\
    mlc-binary & i2c-cont. & 527 & 49.4   & 46.9--53.1     \\
    mlc-binary & stress    & 531 & 70.2   & 66.9--73.0     \\
    \bottomrule
  \end{tabular}
\end{table}

\subsection{Confirmatory tests}

\textbf{H1' (host \textless{} MLC at idle): SUPPORTED.} Host idle median (321.7 $\mu$s) is 359 $\mu$s below MLC bank-switch idle (681.5 $\mu$s), a 2.1$\times$ host speedup. The Hodges-Lehmann shift is $\Delta$HL = $-$359.3 $\mu$s, 95\% bootstrap CI [$-$373.6, $-$352.5] (10,000 iterations); Mann-Whitney p = 1.87 $\times$ 10\textsuperscript{-170}, rejected at the Holm-Bonferroni threshold across the {H1', H2', H3', H6', H7'} family.

\textbf{H2' (host \textless{} MLC under I\textsuperscript{2}C contention): SUPPORTED.} With three concurrent i2c\_hammer processes on bus 7, host median rises to 574.5 $\mu$s and MLC to 1,325.4 $\mu$s; the shift grows to $\Delta$HL = $-$753.2 $\mu$s [$-$760.0, $-$746.7] (2.3$\times$ advantage), p = 2.33 $\times$ 10\textsuperscript{-170}.

\textbf{H3' (MLC degrades more than host): SUPPORTED.} The $\Delta$-of-$\Delta$ contrast is +391.1 $\mu$s [+372.8, +400.0], p = 1.00 $\times$ 10\textsuperscript{-4}. The MLC degrades by +611.7 $\mu$s (+90\%, idle$\rightarrow$contention) versus the host's +249.2 $\mu$s (+78\%): the MLC's three-transaction I\textsuperscript{2}C read protocol amplifies the contention penalty.

\textbf{H5' (CPU stress null for host latency): SUPPORTED.} Host median rises only from 321.7 $\mu$s (idle) to 345.0 $\mu$s (stress); a TOST against the pre-registered $\pm$30 $\mu$s margin gives +23.3 $\mu$s, 90\% CI [+22.7, +23.7] $\subset$ [$-$30, +30]. Equivalence is declared; a 208 Hz polling loop does not contend with stress-ng for CPU time.

\textbf{H6' (CPU stress positive for energy): SUPPORTED.} Mean VDD\_IN (INA3221 via tegrastats) rises from 5,206 mW (idle) to 8,626 mW (stress): +3,420 mW [+3,410, +3,429], exceeding the pre-registered +1,000 mW threshold threefold; H6' is a manipulation check on the CPU-stress condition, not a host-versus-MLC energy comparison (see supplementary S2 for that post-hoc comparison). The power axis distinguishes CPU stress unambiguously where the latency axis (H5') cannot.

\textbf{H7' (MLC stability degrades under contention): FALSIFIED, direction opposite}. The fraction of stimulus windows with exactly one D1 rising edge is 97.22\% (525/540) at idle versus 98.89\% (534/540) under contention, a +1.67 percentage-point \emph{increase} rather than the predicted decrease. Fisher's exact test in the pre-registered direction gives p = 0.9874 (two-sided p = 0.0755). H7' is formally falsified in pre-registration v7.10 \cite{ref4}. I\textsuperscript{2}C contention slows the MLC pipeline (H2', H3') but does not degrade the silicon's classifier reliability.

\subsection{Multimodal latency distributions}

\textbf{Fig.~\ref{fig:latency}} reveals multimodal structure in both MLC pipelines: the mlc/idle distribution is bimodal (mean 866.6 $\mu$s exceeds median 681.5 $\mu$s, p95 1,779.7 $\mu$s), and mlc-binary/idle shows three modes near 60, 240, and 470 $\mu$s. The mlc-binary/i2c-contention distribution has a hidden second mode: median latency is 49.4 $\mu$s, but p95 reaches 246.7 $\mu$s, indicating a secondary high-latency population not captured by the median alone. Both collapse to tighter distributions under contention and stress. We report this as exploratory.

\subsection{MLC decision cadence}

Inter-trial D0 (INT1) gaps (n = 3,086) cluster sharply at integer multiples of T = 706.5 ms, approximately the MLC's 75-sample, 104 Hz window period (0.721 s expected). The quantization is consistent with the MLC updating only on its internal window-cadence clock.

\subsection{Exclusion rates}

No cell exceeded the pre-registered 10\% exclusion ceiling (highest: mlc/idle, 2.78\%). The dominant exclusion category (68 of 90 trials) was multiple D1 edges per window, attributable to the 706.5 ms cadence interacting with stimulus-window boundaries. In the MLC/I\textsuperscript{2}C-contention cell, inclusion (n=532, Table I) and the H7' stability count (n=534, Section~V.A) use related but distinct single-edge criteria: inclusion additionally requires an unambiguous single D0 rising edge before the single D1 edge (Section~III.D), which the H7' criterion does not.

\section{Discussion}

\subsection{The I\textsuperscript{2}C read protocol dominates wire-level latency}

The headline finding, that host classification is 2.1--2.3$\times$ faster than the on-sensor MLC at idle and under I\textsuperscript{2}C contention, is explained by \textbf{Fig.~\ref{fig:latency}(c)}: the mlc-binary pipeline, which performs zero I\textsuperscript{2}C transactions on the decision path (toggling D1 unconditionally on every INT1 edge), reaches a median of 49.4 $\mu$s under contention. The mlc-minus-mlc-binary difference under identical conditions isolates the I\textsuperscript{2}C read overhead at roughly 1,276 $\mu$s under contention (1,325.4 $-$ 49.4) and 476 $\mu$s under stress. Under CPU stress, where the I\textsuperscript{2}C bus itself is uncontended, the host-MLC gap narrows to 1.6$\times$ (345.0 vs 546.1 $\mu$s): the read protocol still costs 476 $\mu$s, but without concurrent bus traffic to amplify it. This cost is not the silicon's classification time; it is the bank-switch read protocol's three-transaction sequence (write FUNC\_CFG\_ACCESS, read MLC0\_SRC, write it back) competing for bus arbitration. A measured SDA/SCL capture of this sequence for one representative trial (confirmatory block b005) is provided in the supplementary material (S4).

\subsection{Implications for safety-critical edge ML}

A naive reading of ``on-sensor inference is faster'' would favor the MLC for low-latency safety-critical loops such as exoskeleton control \cite{ref3}. Our results invert that on the wire-level latency axis: the host reaches its decision 359 $\mu$s earlier at idle, 753 $\mu$s earlier under contention, and is equivalence-null against CPU stress (H5').

The control results sharpen the practical lesson: \textbf{``stress'' is not a single thing.} CPU stress is significant on energy but not latency; I\textsuperscript{2}C contention is significant on latency but not classifier reliability. Bundling both into one ``stress margin'' mis-budgets for each.

\subsection{Multimodal distributions and decision cadence}

First, the MLC pipelines are multimodal at idle (Section~V.B): the mlc/idle p95 of 1,780 $\mu$s is 2.6$\times$ its median, a factor that vanishes under a unimodal-Gaussian assumption. For a ``worst latency observed with probability 1 $-$ $\varepsilon$'' specification, the upper mode, not the median, is the relevant quantity.

Second, for unsynchronized external stimuli, the observed 706.5 ms cadence (Section~V.C) can dominate full \emph{stimulus-to-decision} latency at the system level. Because the MLC fires only on its internal clock boundary, an unsynchronized real-world stimulus waits a uniformly-distributed 0--706.5 ms (mean 353 ms) before the silicon can respond. This is invisible on the D0-to-D1 wire-level axis we measured but is a structural floor; the 1--2 ms wire-level differences this paper characterizes are second-order against it.

\subsection{Limitations}

Results are specific to one platform, sensor IC family, bus protocol (I\textsuperscript{2}C, not SPI), ODR, and MLC configuration; the structural findings should transfer to similar ARM-edge + ST-MEMS combinations (successor parts (LSM6DSV16X, LSM6DSO32X) retain the same bank-switched architecture), but require confirmation. The three-transaction bank-switch cost reflects this implementation's choice to restore the user bank after every MLC0\_SRC read, not an unconditional silicon requirement: an application that caches bank state and defers the switch-back could avoid at least one transaction (not measured here). SPI access would reduce this per-transaction overhead but was not measured, so no speedup is quantified.

A pre-registered RT-scheduling (chrt+taskset) ablation was specified but not activated; pilot data suggest it could roughly halve MLC contention latency, making it the most concrete next step.

\section{Conclusion}

We measured wire-level interrupt-to-decision latency for three pipelines on a Jetson Orin Nano + LSM6DSOX edge platform. The host pipeline is consistently faster; the three-transaction I\textsuperscript{2}C read, not the silicon's classification, dominates.

\end{document}